%% file: albert.tex
\documentstyle[11pt,epsfig]{article}
\newcommand{\bee}{\begin{equation}}
\newcommand{\ee}{\end{equation}}
\newcommand{\beqn}{\begin{eqnarray}}
\newcommand{\eeqn}{\end{eqnarray}}

\newcounter{savefig}
\newcommand{\alphfig}{\setcounter{savefig}{\value{figure}}%
\setcounter{figure}{0}%
\renewcommand{\thefigure}{\mbox{\arabic{savefig}\alph{figure}}}}

\begin{document}
\begin{titlepage}                                                              
\hfill
\hspace*{\fill}
\begin{minipage}[t]{4cm}
DESY--96--168\\
hep-ph/9608401\\
\end{minipage}
\vspace*{2.cm}                                                                 
\begin{center}                                                                 
\begin{LARGE}                                                                  
{\bf The $\gamma^*\gamma^*$ Total Cross Section and the 
BFKL Pomeron at $e^{+}e^{-}$ Colliders}\\
\end{LARGE}                                                                    
\vspace{2.5cm}                                                                   
\begin{Large}
{ 
J. Bartels$^a$, A. De Roeck$^b$ and  H. Lotter$^a$}\\
\end{Large}
\end{center}
\vspace{1.5cm}
$^a$ II.\ Institut f.\ Theoretische Physik, 
Universit\"at Hamburg, Luruper Chaussee 149, \\
D-22761 Hamburg
\footnote
        {Supported by Bundesministerium f\"ur Forschung und
        Technologie, Bonn, Germany under Contract 05\,6HH93P(5) and
        EEC Program "Human Capital and Mobility" through Network
        "Physics at High Energy Colliders" under Contract
        CHRX-CT93-0357 (DG12 COMA).}
\\
$^b$Deutsches Elektronen-Synchrotron, DESY, Notkestr.85, D-22603 Hamburg
\\                                                    
\vspace*{2.cm}                          
\begin{quotation}                                                              
We present a numerical estimate of the $\gamma^* \gamma^*$
total cross section at LEP and at the designed $e^+e^-$
Next Linear Collider (NLC),
based upon the BFKL Pomeron. We find for the linear
collider that the event rate is substantial
provided  electrons scattered under small
angles can be detected, and a measurement of this
cross section provides an excellent test of the BFKL Pomeron.
For LEP, although the number of events is substantially smaller, 
an initial study of this process is feasible. 
\end{quotation}                                                                
\vfill
\noindent
\vspace{1cm}
\end{titlepage}                                                                
\noindent
{\bf 1.} Recently much attention has been given to the BFKL Pomeron 
\cite{BFKL}, in particular in the context of small-$x$
deep inelastic electron proton scattering at HERA. Whereas the use of this
QCD leading logarithmic approximation for the proton structure function is 
affected by serious theoretical difficulties, it has been argued 
\cite{Hot} that the observation of forward jets near the proton beam 
provides a much more reliable test of the BFKL Pomeron. The main reason for
this lies in the fact that the forward jet 
cross section involves only a single 
large momentum scale, namely the transverse momentum of the forward jet 
which is chosen to be
equal or close to the virtuality of the photon. In the structure function 
$F_2$, on the other hand, the BFKL Pomeron feels both the large momentum scale
of the photon mass and the lower factorization scale. In addition, the
diffusion in $\log k_T^2$ always leads to a nonzero contribution of small
transverse momenta where the use of the leading logarithmic approximation
becomes doubtful. A numerical estimate shows that for the forward jets
at HERA \cite{BL} the contribution from this dangerous infrared region is 
reasonably small, whereas in the case of $F_2$ \cite{BLV} the situation is 
much less favorable. As to the experimental situation, a recent analysis of 
HERA data \cite{BDDGW,ADR} on the production of forward jets shows a very 
encouraging agreement between data and the theoretical prediction.\\ \\
In this note we would like to point out that also $e^+e^-$ linear colliders,
in particular the  linear colliders with a rather high luminosity, offer an
excellent opportunity to test the BFKL prediction. The process to be looked at
is the total cross section of $\gamma^* \gamma^*$ scattering. The measurement 
of this cross section requires the double tagging of both outgoing leptons 
close to the forward direction. By varying the energy of the tagged leptons 
it is possible to probe the total cross section of the subprocess
$\gamma^* \gamma^*$ from low energies up to almost the full energy of the 
$e^+ e^-$ collider. For sufficiently large photon virtualities we again have 
a situation with only large momentum scales. In other words, photons with
large virtualities are objects with small transverse sizes, and it is
exactly this situation for which the BFKL approximation should be considered 
to be most reliable. The energy dependence of this cross section, therefore,
should be described by the power law of the BFKL Pomeron. \\
\begin{figure}[!h]
\begin{center}
\input gamma.pstex_t
\caption{Feynman diagram for the process 
$e^+e^- \to e^+e^- + \mbox{anything}$.}
\end{center}
\end{figure}

\noindent
From the theoretical point of view it is clear that we want the photon masses
to be large. On the other hand, because of the photon propagators, the 
$e^+e^-$ cross section for this final state (i.\ e.\ the event rate for the
process under discussion) falls off very rapidly with increasing photon 
masses. Therefore, one cannot afford to have too large photon virtualities. 
As a compromise, we chose the range of $5$ to 
$200 \,\mbox{GeV}^2$ (for 
experimental considerations see further below). As to the energies of the 
$\gamma^*\gamma^*$ subprocess, we can 
in principle go up to almost the full 
collider energies. From the theoretical side, however, 
it is important to estimate the diffusion in the internal 
transverse momenta
$k_T$. In the center of the BFKL ladders (Fig.1), the
distribution in $\log k_T^2$ is given by a Gaussian, with center at
$\log Q^2$ (if we chose, for simplicity, both photon virtualities to be equal),
and with the width growing linearly with 
the square root of the rapidity. 
As soon as the
small-$k_T$ part of the gaussian reaches the confinement region
the BFKL prediction (which is based upon a leading-log calculation) 
becomes unreliable.
Corrections to the BFKL pomeron, in particular those which are expected
to restore unitarity are no longer small.
Qualitatively they are expected to reduce the growth
of the cross section with increasing energy.
Below we will argue that, at least for the 
linear collider, this energy region can be reached. 
In other words, at highest energies for the $\gamma^*\gamma^*$ 
subprocess, a deviation from the power-rise of the BFKL Pomeron might 
become visible.\\ \\ 
{\bf 2.} The theoretical prediction of the cross section is based on  the high 
energy behavior of the diagrams shown in fig.\ 1. Let us first define suitable 
variables. In analogy to DIS kinematics we chose the scaling variables
\beqn
y_1=\frac{q_1 k_2}{k_1k_2}, \,\,\, y_2=\frac{q_2k_1}{k_1k_2} 
\eeqn 
and
\beqn
x_1=\frac{Q_1^2}{2q_1k_2},\,\,\, x_2=\frac{Q_2^2}{2q_2k_1}
\eeqn
where the photon virtualities are, as usual, $Q_i^2=-q_i^2$, $i=1,2$. 
Energies are denoted by $s=(k_1+k_2)^2$ and 
$\hat{s}=(q_1+q_2)^2 \approx sy_1y_2$. With our definitions of the scaling 
variables we have $Q_i^2=sx_iy_i$, $i=1,2$. We consider the limit of large
$Q_1^2$, $Q_2^2$, and $\hat{s}$ with 
\beqn
Q_1^2,\,Q_2^2 \ll \hat{s}.
\eeqn 
The calculation is straightforward and,
neglecting terms of the order of $Q_i^2/\hat{s}$,
leads to the following result:
\beqn
\frac{d \sigma^{e^+e^-} }
{d Q_1^2 d Q_2^2 dy_1 dy_2}
&=& \frac{\alpha^2_{em}}{16 \pi y_1 Q_1^4}
  \frac{\alpha^2_{em}}{16 \pi y_2 Q_2^4}
\int \frac{d \nu}{2 \pi^2} \; \exp\left[\log s 
\frac{y_1 y_2}{\sqrt{Q_1^2 Q_2^2}} \cdot \chi(\nu) \right]
\nonumber \\
& &
\!\!\!\!\!\!\!\!\!\!\!\!\!\!\!\!\!\!\!\!\!\!\!\!\!\!
\!\!\!\!\!\!\!\!\!\!\!\!\!\!\!\!\!\!\!\!\!\!\!\!\!\!\!
\cdot
\left[(1-y_1)W^{(1)}_L(\nu)+\frac{1+(1-y_1)^2}{2}W^{(1)}_T(\nu)\right]
\left[(1-y_2)W^{(2)}_L(-\nu)+\frac{1+(1-y_2)^2}{2}W^{(2)}_T(-\nu)\right] 
\eeqn
with 
$\chi(\nu)= N_c \alpha_s/\pi[2\psi(1)-\psi(1/2+i\nu)-\psi(1/2-i\nu)]$
and we have introduced the invariant functions:
\beqn
W_L^{(i)}(\nu)&=& 
\sum_f q_f^2 \alpha_s \;\pi^2 
\sqrt{2} \; 8  
\frac{\nu^2+\frac{1}{4}}{\nu^2+1} 
\frac{\sinh \pi \nu}{\nu \cosh^2 \pi \nu}
(Q_i^2)^{\frac{1}{2}+i\nu}
\\
W_T^{(i)}(\nu)&=&
\sum_f q_f^2 \alpha_s \;\pi^2 
\sqrt{2} \; 4
\frac{\nu^2+\frac{9}{4}}{\nu^2+1} 
\frac{\sinh \pi \nu}{\nu \cosh^2 \pi \nu}
(Q_i^2)^{\frac{1}{2}+i\nu}
\eeqn
Here $q_f$ is the quark charge, and in our calculations the sum 
over the flavours goes up to three. For the scale of the 
first order strong coupling constant we use $\sqrt{Q_1^2 Q_2^2}$
($\alpha_s(M_Z^2)=0.12$).
In the high energy limit one can use the saddle point approximation near
$\nu=0$ and obtain the following approximation:
\beqn
\frac{d \sigma^{e^+e^-} }
{d Q_1^2 d Q_2^2 dy_1 dy_2}
&=& \frac{\alpha^2_{em}}{16 \pi y_1 Q_1^4}
  \frac{\alpha^2_{em}}{16 \pi y_2 Q_2^4}
\;\exp \left[\log \frac{s}{s_0}  \cdot 
4 \log 2
\frac{N_c \alpha_s}{\pi}  \right]
\;
\frac{e^{-\frac{\log Q_1^2/Q_2^2}{N_c \alpha_s/ \pi 
 56 \zeta(3) \log s/s_0}}}
{\sqrt{N_c \alpha_s /\pi 14 \zeta(3) \log s/s_0}}
\nonumber \\
& & \phantom{xxx} \cdot
\frac{1}{2 \sqrt{\pi}^3}
\;
\sqrt{Q_1^2 Q_2^2}
\left(\sum_f q_f^2 \alpha_s \pi^2 \sqrt{2} \pi\right)^2
\nonumber \\
\nonumber \\
& &
\!\!\!\!\!\!\!\!\!\!\!
\cdot
\left[(1-y_1)W_{L,0}+\frac{1+(1-y_1)^2}{2}W_{T,0}\right]
\left[(1-y_2)W_{L,0}+\frac{1+(1-y_2)^2}{2}W_{T,0}\right] 
\eeqn
where the constants $W_{L,0}, W_{T,0}$ and $s_0$ are:
\beqn
s_0 = \frac{\sqrt{Q_1^2 Q_2^2}}{y_1y_2}
\eeqn
\beqn
W_{L,0}= 2, \,\,\, W_{T,0}=9.
\eeqn 
Before we turn to the discussion of numerical results, let us estimate the
diffusion of internal transverse momenta into the infrared region. In the 
center of the BFKL ladders (fig.\ 1), the width of 
the gaussian distribution of
$\log k_T^2$ is given by \cite{BL}
\beqn
\Delta = \sqrt{<(\log k_T^2 - <\log k_T^2>)^2>} = 
         \sqrt{7 \frac{N_c \alpha_s}{\pi} \zeta(3) \log s/s_0} 
\eeqn
For a typical value $Q_1^2 = Q_2^2=10 \,\mbox{GeV}^2$, and 
$\alpha_s=0.22$ one finds, for the maximal value $Y=\log s/s_0 =10$ at the 
NLC, $\Delta=4.2$ which means that the diffusion reaches down to $k_T^2 
\approx 0.15$ GeV$^2$. For LEP the maximal value for $Y$ is near 6, and
we obtain $\Delta=3.2$ and $k_T^2=0.4 \,\mbox{GeV}^2$. Compared to the 
forward jets at HERA 
\cite{BL} where the corresponding value lies 
above $1 \,\mbox{GeV}^2$, we now have to 
expect a substantially larger contribution from the small-$k_T$ region: this
should lead to a lowering of the BFKL power behavior of the cross section.
In other words, one might be able to see the onset of unitarity corrections 
to the BFKL Pomeron. \\ \\
{\bf 3.} Starting from eqs.\ (4) - (6) we have calculated the differential 
cross section for different values of the logarithm of the subenergy 
$Y=\log s/s_0$.
In order to illustrate
the BFKL power law, we have multiplied the cross section by $y_1y_2$.
In fig.\ 2 we show the results
for both LEP (at the $Z^0$ mass) and the $500$ GeV Next Linear Collider.
On the left hand side we display the $e^+e^-$ cross section
for $Q_1^2=Q_2^2=10 \,\mbox{GeV}^2$. 
The right hand side shows the cross section for 
$Q_1^2=Q_2^2=25 \,\mbox{GeV}^2$. 
\begin{figure}[!h]
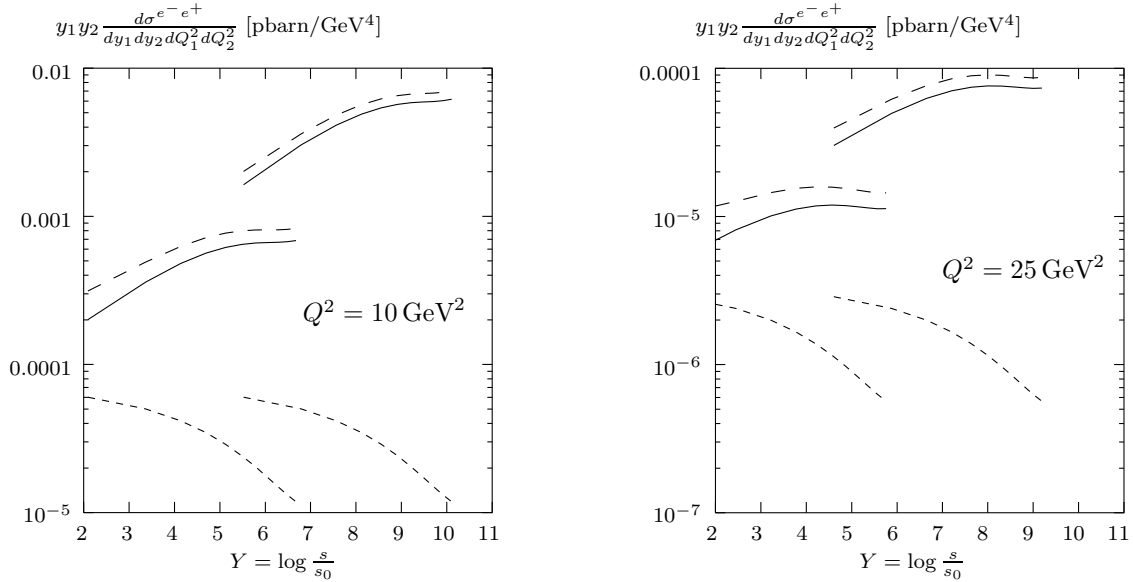

\begin{center}
\input fig2.pstex_t
\caption{
The differential 
$e^+e^-$ cross section, multiplied by $y_1y_2$, as a function
of the rapidity $Y=\log s/s_0$. The full curves denote the exact numerical
calculation based upon eq.\ (4), the dashed curves belong to the analytic
high energy approximation (7), and the dotted lines represent the 
Born approximation (no gluon production between the two quark pairs). 
The three curves on the left in each figure 
belong to LEP, the ones on the right to the Next Linear
Collider NLC. 
We have chosen $y_1=y_2$ and $Q_1^2=Q_2^2 = 10 \,\mbox{GeV}^2$
(left hand figure) resp.\ $Q_1^2=Q_2^2 = 25 \,\mbox{GeV}^2$
(right hand figure).
}
\end{center}
\end{figure}
\\
The full curves represent the ``exact'' results
based upon (4) - (6), the dashed lines show the ``analytic'' prediction of
eq.\ (7), and the dashed dotted lines denote the ``Born'' approximation (no
gluon production between the two fermion pairs). For the exact in the
analytic curves one recognizes the BFKL power-like energy behavior, but there
is some dampening at large rapidity (large $y_i$) due to the photon flux 
factors in (4) and (7). 
In the Born cross section this effect even leads to a decrease
of the $e^+e^-$ cross section at large rapidity $Y$. The BFKL predictions are 
well above the Born curves, up to more than an order of magnitude.
For illustration we show not only the exact but also the analytic 
calculation in order to demonstrate that for our present purposes
the high energy approximation (7) provides a rather good estimate. 
The exact and the analytical calculations agree up
to, typically, a factor less than two. 
Complete agreement between the exact calculation and the approximation
will be reached only at asymptotically high energies.
Comparing the right hand side of fig.\ 2 with the left hand side 
we find that by increasing $Q^2$ from 10 to 25 $\mbox{GeV}^2$
the cross section decreases by two orders of magnitude.
As seen from (4) or (7), the cross section scales with $1/Q^6$, and there is
an additional decrease at larger $Q^2$ due to the $Q^2$-dependence of 
$\alpha_s(Q^2)$. 
\begin{figure}[!h]
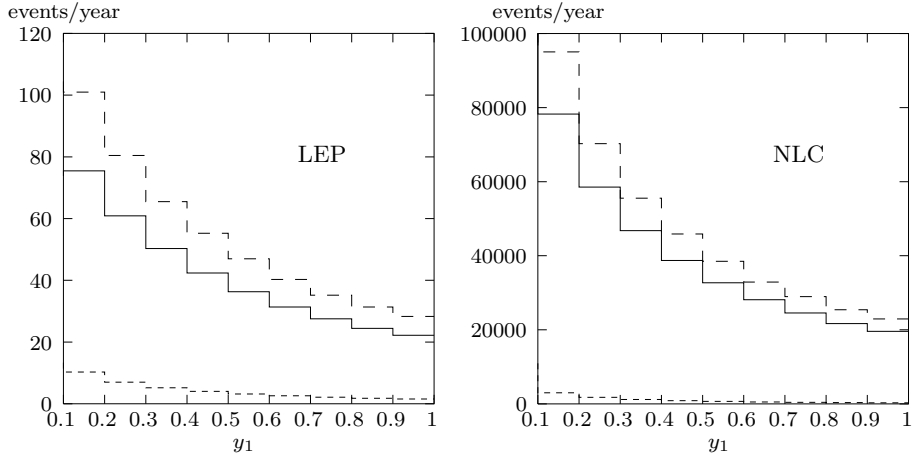

\begin{center}
\input fig3.pstex_t
\caption{
The total number of events per year, as a function of $y_1$. 
The variables
$Q_i^2$ ($5<Q_i^2<200$ GeV$^2$) and $y_2$ ($0.1<y_2<0.9$) are integrated.
The left hand side corresponds to the LEP and the right hand side to 
the NLC situation.
The solid curve shows the results of the exact calculation, the
dashed curve shows the approximation and the dotted curve 
represents the Born result.
}
\end{center}
\end{figure}
Event rates for the NLC and for LEP are shown shown in fig.\ 3, taking 
$3\cdot 10^7$ s/year. Due to the limit in time of data taking periods, 
experiment and accelerator efficiency, this corresponds effectively to several
years of operation of the accelerator.
In $Q_1^2$ and $Q_2^2$ we integrate from $5$ to 
$200 \,\mbox{GeV}^2$, and for the
$y_1$-variable we have chosen 9 bins, as indicated in the figure. The other
$y$-variable is integrated from $0.1$ to $1.0$. The rapidity
of the subprocess $\gamma^* \gamma^*$ is restricted by $\log s/s_0 > 2$.
For the luminosities of LEP and of the
NLC we have used 
$L=10^{31} \mbox{cm}^2 \mbox{s}^{-1}$ and $L=10^{33} \mbox{cm}^2 
\mbox{s}^{-1}$, respectively. 
The event rate calculations are based upon Monte Carlo integration 
of the phase space, and the accuracy is of the order of 5\%.  
The difference of the event rates 
between the NLC and LEP is
a consequence of both the higher luminosity and the higher energy of the NLC.
\\ \\
{\bf 4.} Due to experimental restrictions, however, these event rates
can only give a first impression and not more.
The measurement of the total cross section of $\gamma^* \gamma^*$ scattering
can be made at existing and future $e^+e^-$ colliders
using so called ``double tag'' events. These are events where both 
outgoing leptons are detected and some hadronic activity is observed in the 
central detector.
The $Q^2$ value of the virtual photon emitted from the lepton is
$Q^2= 2E_bE_{tag}(1-\cos\theta_{tag})=
4E_bE_{tag} \sin^2\frac{\theta_{tag}}{2}$, 
with $E_{tag}$ and 
$\theta_{tag}$ the energy and angle of the tagged lepton, 
and $E_b$ the energy of the incident lepton.
The variable $y$ is given by $y=1-(E_{tag}/E_b)\cos^2\frac{\theta_{tag}}{2}$.
Combination of the two relations leads to the convenient equation
\beqn
Q^2=4E_b^2(1-y)\tan^2\frac{\theta_{tag}}{2}
\eeqn
which holds for any of the two incoming leptons.
Experiments at LEP tag electrons down about 60 mrad \cite{delphi,opal}
leading to $Q^2$ values as low as 
to $5-6 \,\mbox{GeV}^2$. In order to reach such 
$Q^2$ values at a NLC with a
CMS energy of 500 GeV, the scattered leptons need to be detected down 
to 10 mrad. For photons of virtuality $20 \,\mbox{GeV}^2$  
angles down to about 15-20 mrad need 
to be covered. 
Present preliminary detector designs intent to tag electrons only 
above 100-150 mrad \cite{NLC}, leading to minimum reachable 
$Q^2$ values in the range of $100-200 \,\mbox{GeV}^2$. 
One of the main problems at small angles is $e^+e^-$
pair production, but preliminary studies indicate that angles down to
20 mrad are within reach \cite{schulte}.
The $y$ values reached in present single tagged analysis at LEP, 
are in the range
$y<0.25$. However using double tagged events the background should 
be kept well under control also for larger $y$ values, and therefore 
values of $y=0.5$ or more, which lead to a large mass system for the 
hadronic final state and, correspondingly, to an extended ladder, are a 
realistic goal. \\ \\
%
\setcounter{figure}{4}
\alphfig
\begin{figure}[!h]
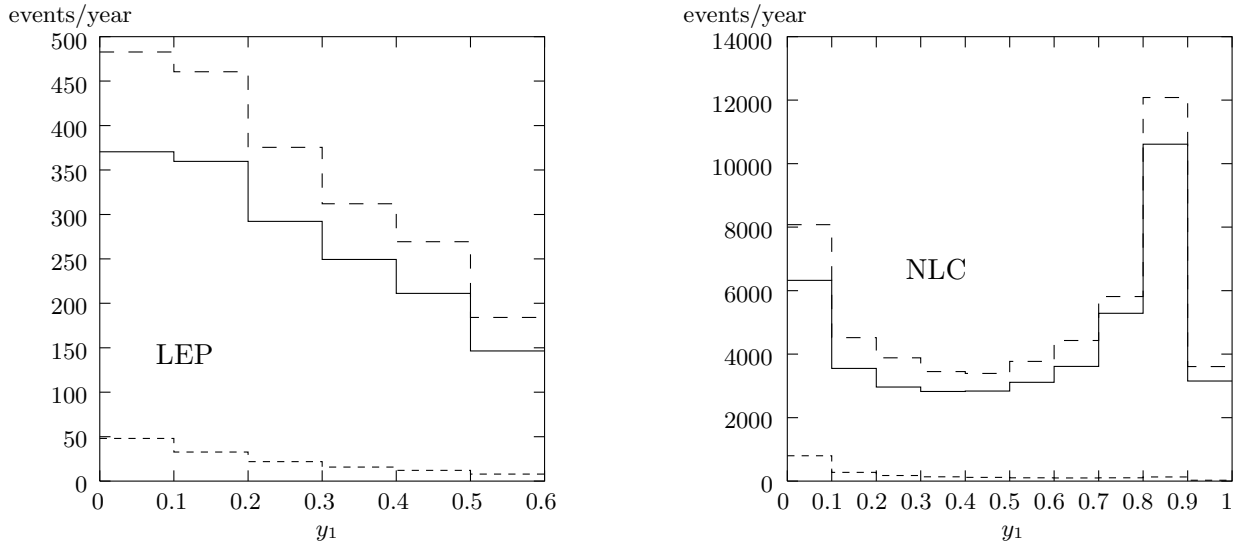

\begin{center}
\input fig44a.pstex_t
\caption{
The total 
number of events per year ($3\cdot 10^7$ s)
with detector cuts in angle taken into account.
We have chosen $E_{tag}>20$ GeV, $\log s/s_0>2$, $2.5 <Q_i^2 <200$ GeV$^2$.
The acceptance cut is $\theta_{tag}> 20 \,\mbox{mrad}$. 
The $y$-binning is the same as in fig. 3 and we have LEP on the left and 
NLC on the right hand side.
}
\end{center}
\end{figure}
\begin{figure}[!h]
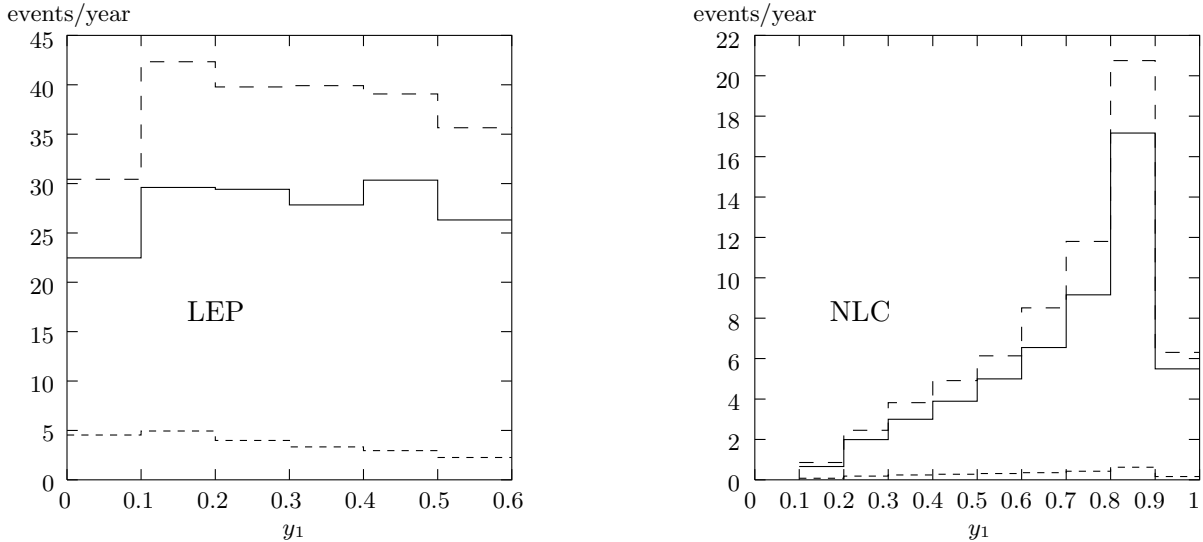

\begin{center}
\input fig44b.pstex_t
\caption{
The same as in fig.\ 4a but with acceptance cut 
$\theta_{tag}> 60 \,\mbox{mrad}$.
}
\end{center}
\end{figure}
In order to incorporate these conditions into our calculation of event rates,
we have repeated the previous calculations with more realistic kinematical
cuts. For both the NLC and LEP we impose the constraints 
$\theta_{tag}>20$ mrad (fig.\ 4a) and $\theta_{tag}>60$ mrad
(fig.\ 4b), and
we require $E_{tag}>20$ GeV (i.\ e.\ approximately 
$y_i<0.9$ for the NLC and $y_i<0.6$ for LEP). 
Further restrictions are $2.5<Q_i^2<200 \,\mbox{GeV}^2$ and
$\log s/s_0>2$.
For the NLC case, in particular for smaller $y_1$, the 
event rate is 
substantially lower than in fig.\ 3, indicating the importance of the detector
angle cut. Clearly, $\theta_{tag}=20$ mrad looks highly desirable for both 
machines. For LEP, the comparison with the NLC now looks more favorable than
in fig.\ 3. If $\theta_{tag}$ can be taken down to 20 mrad, the NLC rate is
clearly substantially larger than the LEP rate. For $\theta_{tag}=60$ mrad, 
however, the $Q^2$ values at the NLC are considerably larger than those at
LEP (cf.\ eq.\ (11)), and the small-$y$ region 
in the right hand diagram 
in fig.\ 4b has very few events.
This explains why, despite the higher energies and the larger luminosity of 
the NLC, LEP is doing better than the NLC. 
%
\\ \\
Finally, we have computed a few integrated event rates. The results are shown 
in Table 1 for the NLC, LEP-90 and LEP-180.  
%
\begin{table}[!h]
\begin{center}
\begin{tabular}{|c|c|c|c|c|} \hline
exact & analytic & Born & $\theta_{tag,\,\,min}$ & \\ 
\hline  \hline
29500 & 36400 & 495 & 20 mrad & NLC\\
51 & 68 & 2 & 60 mrad &  \\
1 & 1 & 0 & 100 mrad & \\ 
\hline  \hline
940 & 1150 & 47 & 20 mrad  & LEP-90\\
120 & 158 &13 & 60 mrad  & \\
11 & 15 & 2 & 100 mrad &  \\
\hline
3250 & 3870 & 53 & 20 mrad  & LEP-180\\
 750 & 880  & 13 & 30 mrad & \\
24 & 30 & 3 & 60 mrad  & \\
2 & 3 & 0 & 100 mrad &  \\ \hline
\end{tabular} 
\caption{
Total event rates per year for NLC, LEP I and LEP II.
}
\end{center}
\end{table}
%
We integrated $2.5<Q_i^2<200
\,\mbox{GeV}^2$, 
$0.1<y_i<1.0$, with the
constraints $E_{tag}>20 \,\mbox{GeV}$, 
$\log s/s_0>2$, and for the detector angles
we have chosen $\theta_{tag}>20$ mrad, $60$ mrad, and $100$ mrad.
For the smallest angles 
the ratio of BFKL to Born cross section is larger than a factor of
75 for the NLC, while it is reduced 
to 25 at LEP-90. 
For angles above 60 mrad for which 
single tag measurements at LEP exist,
the BFKL cross section is by a 
factor of 10 larger than the Born cross section.
This factor is larger at the NLC,
but the limitation in $Q^2$ strongly affects the rate as discussed above. 
The total number of events produced in the four LEP experiments 
together at LEP-90, taking a total collected integrated luminosity
of 150 pb$^{-1}$/experiment is twice the number of events given in Table 1.
Reversely, the LEP-180 numbers correspond to a total number of events
collected by the experiments if 75 pb$^{-1}$ will be delivered by the machine.
The experiments at LEP-180 are foreseen to measure already now 
down to angles about 30 mrad. 
Table 1 shows that the event rate for this angle
is reasonable.
We conclude that, mainly because of the higher luminosity of the NLC,
such a machine offers an excellent possibility 
to observe the BFKL Pomeron, provided the detectors can be improved to reach
small angles 
down to about $20$ mrad. For LEP the total event rate looks less encouraging,
but it is still worthwhile to pursue further studies in this direction.
\\ \\
{\bf 5.} In summary, we have estimated the total cross section of
$\gamma^* \gamma^*$ scattering at both LEP and the  NLC.
At high energies, this subprocess is dominated by the BFKL Pomeron and 
therefore provides an ideal test of this QCD calculation. A rough estimate of 
the diffusion in $\log k_T^2$ shows that, at the high energy tail of this 
subprocess, corrections to the BFKL become non-negligible, and hence 
deviations from the BFKL power law may become visible.\\ \\
For a realistic estimate of the number of observable events we find a strong
dependence on detector restrictions. In order to have a sufficiently large
number of events it is necessary to tag both leptons close to the beam 
direction; a desirable angle would be 20 mrad, and energies of the
tagged leptons should go down to about $20 \,\mbox{GeV}$. 
In this region the NLC
provides an excellent possibility for testing the BFKL Pomeron. For LEP
(in particular at $90$ GeV lepton energies) the number of events is still 
sufficiently large to justify a dedicated search. For larger angles 
(e.g. $60$ mrad) the NLC loses in the number of events, faster than LEP. 
\\ \\
{\bf Acknowledgements:} We wish to thank Peter Zerwas, Daniel Schulte 
and Richard Nisius for 
very helpful discussions.\\ \\
{\bf Note added:} 
Results similar to those contained in this paper
have been 
obtained independently
by S.\ Brodsky,  
F.\ Hautmann and  
D.\ Soper \cite{haut}.
%

\end{document}

%% file: gamma.pstex_t
\begin{picture}(0,0)%
\epsfig{file=gamma.pstex}%
\end{picture}%
\setlength{\unitlength}{0.00065600in}%
\begingroup\makeatletter\ifx\SetFigFont\undefined
\def\x#1#2#3#4#5#6#7\relax{\def\x{#1#2#3#4#5#6}}%
\expandafter\x\fmtname xxxxxx\relax \def\y{splain}%
\ifx\x\y   
\gdef\SetFigFont#1#2#3{%
  \ifnum #1<17\tiny\else \ifnum #1<20\small\else
  \ifnum #1<24\normalsize\else \ifnum #1<29\large\else
  \ifnum #1<34\Large\else \ifnum #1<41\LARGE\else
     \huge\fi\fi\fi\fi\fi\fi
  \csname #3\endcsname}%
\else
\gdef\SetFigFont#1#2#3{\begingroup
  \count@#1\relax \ifnum 25<\count@\count@25\fi
  \def\x{\endgroup\@setsize\SetFigFont{#2pt}}%
  \expandafter\x
    \csname \romannumeral\the\count@ pt\expandafter\endcsname
    \csname @\romannumeral\the\count@ pt\endcsname
  \csname #3\endcsname}%
\fi
\fi\endgroup
\begin{picture}(3375,5520)(2971,-6709)
\put(6346,-2581){\makebox(0,0)[lb]{\smash{\SetFigFont{9}{10.8}{rm}$q$}}}
\put(6346,-3031){\makebox(0,0)[lb]{\smash{\SetFigFont{9}{10.8}{rm}$\bar{q}$}}}
\put(6346,-4831){\makebox(0,0)[lb]{\smash{\SetFigFont{9}{10.8}{rm}$q$}}}
\put(6346,-5281){\makebox(0,0)[lb]{\smash{\SetFigFont{9}{10.8}{rm}$\bar{q}$}}}
\put(4771,-5866){\makebox(0,0)[lb]{\smash{\SetFigFont{9}{10.8}{rm}$\gamma^*$}}}
\put(4771,-2041){\makebox(0,0)[lb]{\smash{\SetFigFont{9}{10.8}{rm}$\gamma^*$}}}
\put(4501,-6676){\makebox(0,0)[lb]{\smash{\SetFigFont{9}{10.8}{rm}$e'^+$}}}
\put(4546,-1681){\makebox(0,0)[lb]{\smash{\SetFigFont{9}{10.8}{rm}$k_1'$}}}
\put(4501,-6181){\makebox(0,0)[lb]{\smash{\SetFigFont{9}{10.8}{rm}$k_2'$}}}
\put(4186,-2401){\makebox(0,0)[lb]{\smash{\SetFigFont{9}{10.8}{rm}$q_1$}}}
\put(4186,-5461){\makebox(0,0)[lb]{\smash{\SetFigFont{9}{10.8}{rm}$q_2$}}}
\put(2971,-5956){\makebox(0,0)[lb]{\smash{\SetFigFont{9}{10.8}{rm}$k_2$}}}
\put(4501,-1321){\makebox(0,0)[lb]{\smash{\SetFigFont{9}{10.8}{rm}$e'^-$}}}
\put(3466,-6226){\makebox(0,0)[lb]{\smash{\SetFigFont{9}{10.8}{rm}$e^+$}}}
\put(3466,-1636){\makebox(0,0)[lb]{\smash{\SetFigFont{9}{10.8}{rm}$e^-$}}}
\put(3016,-1861){\makebox(0,0)[lb]{\smash{\SetFigFont{9}{10.8}{rm}$k_1$}}}
\end{picture}

%% file: fig2.pstex_t
\begin{picture}(0,0)%
\epsfig{file=fig2.pstex}%
\end{picture}%
\setlength{\unitlength}{0.00075000in}%
\begingroup\makeatletter\ifx\SetFigFont\undefined
\def\x#1#2#3#4#5#6#7\relax{\def\x{#1#2#3#4#5#6}}%
\expandafter\x\fmtname xxxxxx\relax \def\y{splain}%
\ifx\x\y   
\gdef\SetFigFont#1#2#3{%
  \ifnum #1<17\tiny\else \ifnum #1<20\small\else
  \ifnum #1<24\normalsize\else \ifnum #1<29\large\else
  \ifnum #1<34\Large\else \ifnum #1<41\LARGE\else
     \huge\fi\fi\fi\fi\fi\fi
  \csname #3\endcsname}%
\else
\gdef\SetFigFont#1#2#3{\begingroup
  \count@#1\relax \ifnum 25<\count@\count@25\fi
  \def\x{\endgroup\@setsize\SetFigFont{#2pt}}%
  \expandafter\x
    \csname \romannumeral\the\count@ pt\expandafter\endcsname
    \csname @\romannumeral\the\count@ pt\endcsname
  \csname #3\endcsname}%
\fi
\fi\endgroup
\begin{picture}(10243,3901)(67,-3841)
\put(6032,-3620){\makebox(0,0)[b]{\smash{\SetFigFont{8}{9.6}{rm}4}}}
\put(6349,-3620){\makebox(0,0)[b]{\smash{\SetFigFont{8}{9.6}{rm}5}}}
\put(6666,-3620){\makebox(0,0)[b]{\smash{\SetFigFont{8}{9.6}{rm}6}}}
\put(6982,-3620){\makebox(0,0)[b]{\smash{\SetFigFont{8}{9.6}{rm}7}}}
\put(7298,-3620){\makebox(0,0)[b]{\smash{\SetFigFont{8}{9.6}{rm}8}}}
\put(5715,-3620){\makebox(0,0)[b]{\smash{\SetFigFont{8}{9.6}{rm}3}}}
\put(3206,-3619){\makebox(0,0)[b]{\smash{\SetFigFont{8}{9.6}{rm}9}}}
\put(3523,-3619){\makebox(0,0)[b]{\smash{\SetFigFont{8}{9.6}{rm}10}}}
\put(3840,-3619){\makebox(0,0)[b]{\smash{\SetFigFont{8}{9.6}{rm}11}}}
\put(5330,-392){\makebox(0,0)[rb]{\smash{\SetFigFont{8}{9.6}{rm}0.0001}}}
\put(5398,-3620){\makebox(0,0)[b]{\smash{\SetFigFont{8}{9.6}{rm}2}}}
\put(7615,-3620){\makebox(0,0)[b]{\smash{\SetFigFont{8}{9.6}{rm}9}}}
\put(5330,-1427){\makebox(0,0)[rb]{\smash{\SetFigFont{8}{9.6}{rm}$10^{-5}$}}}
\put(5330,-2461){\makebox(0,0)[rb]{\smash{\SetFigFont{8}{9.6}{rm}$10^{-6}$}}}
\put(5330,-3496){\makebox(0,0)[rb]{\smash{\SetFigFont{8}{9.6}{rm}$10^{-7}$}}}
\put(6428,-62){\makebox(0,0)[b]{\smash{\SetFigFont{8}{9.6}{rm}$y_1 y_2 \frac{d \sigma^{e^-e^+}}{dy_1 dy_2 dQ_1^2 dQ_2^2} \; [\mbox{pbarn}/\mbox{GeV}^4]$}}}
\put(1950,-61){\makebox(0,0)[b]{\smash{\SetFigFont{8}{9.6}{rm}$y_1 y_2 \frac{d \sigma^{e^-e^+}}{dy_1 dy_2 dQ_1^2 dQ_2^2} \; [\mbox{pbarn}/\mbox{GeV}^4]$}}}
\put(921,-3495){\makebox(0,0)[rb]{\smash{\SetFigFont{8}{9.6}{rm}$10^{-5}$}}}
\put(7932,-3620){\makebox(0,0)[b]{\smash{\SetFigFont{8}{9.6}{rm}10}}}
\put(8249,-3620){\makebox(0,0)[b]{\smash{\SetFigFont{8}{9.6}{rm}11}}}
\put(2399,-3811){\makebox(0,0)[b]{\smash{\SetFigFont{8}{9.6}{rm}$Y = \log \frac{s}{s_0}$}}}
\put(6807,-3811){\makebox(0,0)[b]{\smash{\SetFigFont{8}{9.6}{rm}$Y = \log \frac{s}{s_0}$}}}
\put(2537,-2086){\makebox(0,0)[lb]{\smash{\SetFigFont{10}{12.0}{rm}$Q^2 = 10 \,\mbox{GeV}^2$}}}
\put(7014,-1786){\makebox(0,0)[lb]{\smash{\SetFigFont{10}{12.0}{rm}$Q^2 = 25 \,\mbox{GeV}^2$}}}
\put(2889,-3619){\makebox(0,0)[b]{\smash{\SetFigFont{8}{9.6}{rm}8}}}
\put(1306,-3619){\makebox(0,0)[b]{\smash{\SetFigFont{8}{9.6}{rm}3}}}
\put(1623,-3619){\makebox(0,0)[b]{\smash{\SetFigFont{8}{9.6}{rm}4}}}
\put(1940,-3619){\makebox(0,0)[b]{\smash{\SetFigFont{8}{9.6}{rm}5}}}
\put(2257,-3619){\makebox(0,0)[b]{\smash{\SetFigFont{8}{9.6}{rm}6}}}
\put(2572,-3619){\makebox(0,0)[b]{\smash{\SetFigFont{8}{9.6}{rm}7}}}
\put(989,-3619){\makebox(0,0)[b]{\smash{\SetFigFont{8}{9.6}{rm}2}}}
\put(921,-2460){\makebox(0,0)[rb]{\smash{\SetFigFont{8}{9.6}{rm}0.0001}}}
\put(921,-1426){\makebox(0,0)[rb]{\smash{\SetFigFont{8}{9.6}{rm}0.001}}}
\put(921,-391){\makebox(0,0)[rb]{\smash{\SetFigFont{8}{9.6}{rm}0.01}}}
\end{picture}

%% file: fig3.pstex_t
\begin{picture}(0,0)%
\epsfig{file=fig3.pstex}%
\end{picture}%
\setlength{\unitlength}{0.00062500in}%
\begingroup\makeatletter\ifx\SetFigFont\undefined
\def\x#1#2#3#4#5#6#7\relax{\def\x{#1#2#3#4#5#6}}%
\expandafter\x\fmtname xxxxxx\relax \def\y{splain}%
\ifx\x\y   
\gdef\SetFigFont#1#2#3{%
  \ifnum #1<17\tiny\else \ifnum #1<20\small\else
  \ifnum #1<24\normalsize\else \ifnum #1<29\large\else
  \ifnum #1<34\Large\else \ifnum #1<41\LARGE\else
     \huge\fi\fi\fi\fi\fi\fi
  \csname #3\endcsname}%
\else
\gdef\SetFigFont#1#2#3{\begingroup
  \count@#1\relax \ifnum 25<\count@\count@25\fi
  \def\x{\endgroup\@setsize\SetFigFont{#2pt}}%
  \expandafter\x
    \csname \romannumeral\the\count@ pt\expandafter\endcsname
    \csname @\romannumeral\the\count@ pt\endcsname
  \csname #3\endcsname}%
\fi
\fi\endgroup
\begin{picture}(7437,3742)(622,-3094)
\put(870,-1716){\makebox(0,0)[rb]{\smash{\SetFigFont{8}{9.6}{rm}40}}}
\put(870,-2234){\makebox(0,0)[rb]{\smash{\SetFigFont{8}{9.6}{rm}20}}}
\put(870,-2751){\makebox(0,0)[rb]{\smash{\SetFigFont{8}{9.6}{rm}0}}}
\put(870,-1199){\makebox(0,0)[rb]{\smash{\SetFigFont{8}{9.6}{rm}60}}}
\put(1634,-2875){\makebox(0,0)[b]{\smash{\SetFigFont{8}{9.6}{rm}0.3}}}
\put(1289,-2875){\makebox(0,0)[b]{\smash{\SetFigFont{8}{9.6}{rm}0.2}}}
\put(944,-2875){\makebox(0,0)[b]{\smash{\SetFigFont{8}{9.6}{rm}0.1}}}
\put(870,353){\makebox(0,0)[rb]{\smash{\SetFigFont{8}{9.6}{rm}120}}}
\put(870,-164){\makebox(0,0)[rb]{\smash{\SetFigFont{8}{9.6}{rm}100}}}
\put(870,-682){\makebox(0,0)[rb]{\smash{\SetFigFont{8}{9.6}{rm}80}}}
\put(1979,-2875){\makebox(0,0)[b]{\smash{\SetFigFont{8}{9.6}{rm}0.4}}}
\put(7331,-2876){\makebox(0,0)[b]{\smash{\SetFigFont{8}{9.6}{rm}0.8}}}
\put(6986,-2876){\makebox(0,0)[b]{\smash{\SetFigFont{8}{9.6}{rm}0.7}}}
\put(6641,-2876){\makebox(0,0)[b]{\smash{\SetFigFont{8}{9.6}{rm}0.6}}}
\put(6297,-2876){\makebox(0,0)[b]{\smash{\SetFigFont{8}{9.6}{rm}0.5}}}
\put(5952,-2876){\makebox(0,0)[b]{\smash{\SetFigFont{8}{9.6}{rm}0.4}}}
\put(5607,-2876){\makebox(0,0)[b]{\smash{\SetFigFont{8}{9.6}{rm}0.3}}}
\put(7676,-2876){\makebox(0,0)[b]{\smash{\SetFigFont{8}{9.6}{rm}0.9}}}
\put(6901,-661){\makebox(0,0)[lb]{\smash{\SetFigFont{9}{10.8}{rm}NLC}}}
\put(2926,-661){\makebox(0,0)[lb]{\smash{\SetFigFont{9}{10.8}{rm}LEP}}}
\put(6451,-3061){\makebox(0,0)[b]{\smash{\SetFigFont{8}{9.6}{rm}$y_1$}}}
\put(2476,-3061){\makebox(0,0)[b]{\smash{\SetFigFont{8}{9.6}{rm}$y_1$}}}
\put(5027,538){\makebox(0,0)[b]{\smash{\SetFigFont{8}{9.6}{rm}events/year}}}
\put(8021,-2876){\makebox(0,0)[b]{\smash{\SetFigFont{8}{9.6}{rm}1}}}
\put(5262,-2876){\makebox(0,0)[b]{\smash{\SetFigFont{8}{9.6}{rm}0.2}}}
\put(4048,-2875){\makebox(0,0)[b]{\smash{\SetFigFont{8}{9.6}{rm}1}}}
\put(3703,-2875){\makebox(0,0)[b]{\smash{\SetFigFont{8}{9.6}{rm}0.9}}}
\put(3358,-2875){\makebox(0,0)[b]{\smash{\SetFigFont{8}{9.6}{rm}0.8}}}
\put(3013,-2875){\makebox(0,0)[b]{\smash{\SetFigFont{8}{9.6}{rm}0.7}}}
\put(2668,-2875){\makebox(0,0)[b]{\smash{\SetFigFont{8}{9.6}{rm}0.6}}}
\put(2324,-2875){\makebox(0,0)[b]{\smash{\SetFigFont{8}{9.6}{rm}0.5}}}
\put(976,539){\makebox(0,0)[b]{\smash{\SetFigFont{8}{9.6}{rm}events/year}}}
\put(4917,-2876){\makebox(0,0)[b]{\smash{\SetFigFont{8}{9.6}{rm}0.1}}}
\put(4843,352){\makebox(0,0)[rb]{\smash{\SetFigFont{8}{9.6}{rm}100000}}}
\put(4843,-269){\makebox(0,0)[rb]{\smash{\SetFigFont{8}{9.6}{rm}80000}}}
\put(4843,-890){\makebox(0,0)[rb]{\smash{\SetFigFont{8}{9.6}{rm}60000}}}
\put(4843,-1510){\makebox(0,0)[rb]{\smash{\SetFigFont{8}{9.6}{rm}40000}}}
\put(4843,-2131){\makebox(0,0)[rb]{\smash{\SetFigFont{8}{9.6}{rm}20000}}}
\put(4843,-2752){\makebox(0,0)[rb]{\smash{\SetFigFont{8}{9.6}{rm}0}}}
\end{picture}

%% file: fig44a.pstex_t
\begin{picture}(0,0)%
\epsfig{file=fig44a.pstex}%
\end{picture}%
\setlength{\unitlength}{0.00075000in}%
\begingroup\makeatletter\ifx\SetFigFont\undefined
\def\x#1#2#3#4#5#6#7\relax{\def\x{#1#2#3#4#5#6}}%
\expandafter\x\fmtname xxxxxx\relax \def\y{splain}%
\ifx\x\y   
\gdef\SetFigFont#1#2#3{%
  \ifnum #1<17\tiny\else \ifnum #1<20\small\else
  \ifnum #1<24\normalsize\else \ifnum #1<29\large\else
  \ifnum #1<34\Large\else \ifnum #1<41\LARGE\else
     \huge\fi\fi\fi\fi\fi\fi
  \csname #3\endcsname}%
\else
\gdef\SetFigFont#1#2#3{\begingroup
  \count@#1\relax \ifnum 25<\count@\count@25\fi
  \def\x{\endgroup\@setsize\SetFigFont{#2pt}}%
  \expandafter\x
    \csname \romannumeral\the\count@ pt\expandafter\endcsname
    \csname @\romannumeral\the\count@ pt\endcsname
  \csname #3\endcsname}%
\fi
\fi\endgroup
\begin{picture}(8488,3742)(697,-3094)
\put(1171,-2131){\makebox(0,0)[rb]{\smash{\SetFigFont{9}{10.8}{rm}100}}}
\put(1171,-2442){\makebox(0,0)[rb]{\smash{\SetFigFont{9}{10.8}{rm}50}}}
\put(1171,-2752){\makebox(0,0)[rb]{\smash{\SetFigFont{9}{10.8}{rm}0}}}
\put(1171,-1821){\makebox(0,0)[rb]{\smash{\SetFigFont{9}{10.8}{rm}150}}}
\put(1171,352){\makebox(0,0)[rb]{\smash{\SetFigFont{9}{10.8}{rm}500}}}
\put(1171, 42){\makebox(0,0)[rb]{\smash{\SetFigFont{9}{10.8}{rm}450}}}
\put(1171,-269){\makebox(0,0)[rb]{\smash{\SetFigFont{9}{10.8}{rm}400}}}
\put(1171,-579){\makebox(0,0)[rb]{\smash{\SetFigFont{9}{10.8}{rm}350}}}
\put(1171,-890){\makebox(0,0)[rb]{\smash{\SetFigFont{9}{10.8}{rm}300}}}
\put(1171,-1200){\makebox(0,0)[rb]{\smash{\SetFigFont{9}{10.8}{rm}250}}}
\put(1171,-1510){\makebox(0,0)[rb]{\smash{\SetFigFont{9}{10.8}{rm}200}}}
\put(1245,-2876){\makebox(0,0)[b]{\smash{\SetFigFont{9}{10.8}{rm}0}}}
\put(8526,-2876){\makebox(0,0)[b]{\smash{\SetFigFont{9}{10.8}{rm}0.8}}}
\put(8216,-2876){\makebox(0,0)[b]{\smash{\SetFigFont{9}{10.8}{rm}0.7}}}
\put(7905,-2876){\makebox(0,0)[b]{\smash{\SetFigFont{9}{10.8}{rm}0.6}}}
\put(7595,-2876){\makebox(0,0)[b]{\smash{\SetFigFont{9}{10.8}{rm}0.5}}}
\put(7285,-2876){\makebox(0,0)[b]{\smash{\SetFigFont{9}{10.8}{rm}0.4}}}
\put(6974,-2876){\makebox(0,0)[b]{\smash{\SetFigFont{9}{10.8}{rm}0.3}}}
\put(6664,-2876){\makebox(0,0)[b]{\smash{\SetFigFont{9}{10.8}{rm}0.2}}}
\put(8837,-2876){\makebox(0,0)[b]{\smash{\SetFigFont{9}{10.8}{rm}0.9}}}
\put(5776,539){\makebox(0,0)[b]{\smash{\SetFigFont{9}{10.8}{rm}events/year}}}
\put(1051,539){\makebox(0,0)[b]{\smash{\SetFigFont{9}{10.8}{rm}events/year}}}
\put(6901,-1261){\makebox(0,0)[lb]{\smash{\SetFigFont{11}{13.2}{rm}NLC}}}
\put(1651,-1861){\makebox(0,0)[lb]{\smash{\SetFigFont{11}{13.2}{rm}LEP}}}
\put(7651,-3061){\makebox(0,0)[b]{\smash{\SetFigFont{9}{10.8}{rm}$y_1$}}}
\put(2851,-3061){\makebox(0,0)[b]{\smash{\SetFigFont{9}{10.8}{rm}$y_1$}}}
\put(9147,-2876){\makebox(0,0)[b]{\smash{\SetFigFont{9}{10.8}{rm}1}}}
\put(6353,-2876){\makebox(0,0)[b]{\smash{\SetFigFont{9}{10.8}{rm}0.1}}}
\put(5969,-2752){\makebox(0,0)[rb]{\smash{\SetFigFont{9}{10.8}{rm}0}}}
\put(4349,-2876){\makebox(0,0)[b]{\smash{\SetFigFont{9}{10.8}{rm}0.6}}}
\put(3832,-2876){\makebox(0,0)[b]{\smash{\SetFigFont{9}{10.8}{rm}0.5}}}
\put(3314,-2876){\makebox(0,0)[b]{\smash{\SetFigFont{9}{10.8}{rm}0.4}}}
\put(2797,-2876){\makebox(0,0)[b]{\smash{\SetFigFont{9}{10.8}{rm}0.3}}}
\put(2280,-2876){\makebox(0,0)[b]{\smash{\SetFigFont{9}{10.8}{rm}0.2}}}
\put(1762,-2876){\makebox(0,0)[b]{\smash{\SetFigFont{9}{10.8}{rm}0.1}}}
\put(5969,-2309){\makebox(0,0)[rb]{\smash{\SetFigFont{9}{10.8}{rm}2000}}}
\put(6043,-2876){\makebox(0,0)[b]{\smash{\SetFigFont{9}{10.8}{rm}0}}}
\put(5969,352){\makebox(0,0)[rb]{\smash{\SetFigFont{9}{10.8}{rm}14000}}}
\put(5969,-91){\makebox(0,0)[rb]{\smash{\SetFigFont{9}{10.8}{rm}12000}}}
\put(5969,-535){\makebox(0,0)[rb]{\smash{\SetFigFont{9}{10.8}{rm}10000}}}
\put(5969,-978){\makebox(0,0)[rb]{\smash{\SetFigFont{9}{10.8}{rm}8000}}}
\put(5969,-1422){\makebox(0,0)[rb]{\smash{\SetFigFont{9}{10.8}{rm}6000}}}
\put(5969,-1865){\makebox(0,0)[rb]{\smash{\SetFigFont{9}{10.8}{rm}4000}}}
\end{picture}

%% file: fig44b.pstex_t
\begin{picture}(0,0)%
\epsfig{file=fig44b.pstex}%
\end{picture}%
\setlength{\unitlength}{0.00075000in}%
\begingroup\makeatletter\ifx\SetFigFont\undefined
\def\x#1#2#3#4#5#6#7\relax{\def\x{#1#2#3#4#5#6}}%
\expandafter\x\fmtname xxxxxx\relax \def\y{splain}%
\ifx\x\y   
\gdef\SetFigFont#1#2#3{%
  \ifnum #1<17\tiny\else \ifnum #1<20\small\else
  \ifnum #1<24\normalsize\else \ifnum #1<29\large\else
  \ifnum #1<34\Large\else \ifnum #1<41\LARGE\else
     \huge\fi\fi\fi\fi\fi\fi
  \csname #3\endcsname}%
\else
\gdef\SetFigFont#1#2#3{\begingroup
  \count@#1\relax \ifnum 25<\count@\count@25\fi
  \def\x{\endgroup\@setsize\SetFigFont{#2pt}}%
  \expandafter\x
    \csname \romannumeral\the\count@ pt\expandafter\endcsname
    \csname @\romannumeral\the\count@ pt\endcsname
  \csname #3\endcsname}%
\fi
\fi\endgroup
\begin{picture}(8266,3742)(922,-3094)
\put(1172,-1717){\makebox(0,0)[rb]{\smash{\SetFigFont{9}{10.8}{rm}15}}}
\put(1172,-2062){\makebox(0,0)[rb]{\smash{\SetFigFont{9}{10.8}{rm}10}}}
\put(1172,-2407){\makebox(0,0)[rb]{\smash{\SetFigFont{9}{10.8}{rm}5}}}
\put(1172,-2752){\makebox(0,0)[rb]{\smash{\SetFigFont{9}{10.8}{rm}0}}}
\put(1172,-1372){\makebox(0,0)[rb]{\smash{\SetFigFont{9}{10.8}{rm}20}}}
\put(1763,-2876){\makebox(0,0)[b]{\smash{\SetFigFont{9}{10.8}{rm}0.1}}}
\put(1246,-2876){\makebox(0,0)[b]{\smash{\SetFigFont{9}{10.8}{rm}0}}}
\put(1172,352){\makebox(0,0)[rb]{\smash{\SetFigFont{9}{10.8}{rm}45}}}
\put(1172,  7){\makebox(0,0)[rb]{\smash{\SetFigFont{9}{10.8}{rm}40}}}
\put(1172,-338){\makebox(0,0)[rb]{\smash{\SetFigFont{9}{10.8}{rm}35}}}
\put(1172,-683){\makebox(0,0)[rb]{\smash{\SetFigFont{9}{10.8}{rm}30}}}
\put(1172,-1028){\makebox(0,0)[rb]{\smash{\SetFigFont{9}{10.8}{rm}25}}}
\put(2281,-2876){\makebox(0,0)[b]{\smash{\SetFigFont{9}{10.8}{rm}0.2}}}
\put(8219,-2876){\makebox(0,0)[b]{\smash{\SetFigFont{9}{10.8}{rm}0.7}}}
\put(7908,-2876){\makebox(0,0)[b]{\smash{\SetFigFont{9}{10.8}{rm}0.6}}}
\put(7598,-2876){\makebox(0,0)[b]{\smash{\SetFigFont{9}{10.8}{rm}0.5}}}
\put(7288,-2876){\makebox(0,0)[b]{\smash{\SetFigFont{9}{10.8}{rm}0.4}}}
\put(6977,-2876){\makebox(0,0)[b]{\smash{\SetFigFont{9}{10.8}{rm}0.3}}}
\put(6667,-2876){\makebox(0,0)[b]{\smash{\SetFigFont{9}{10.8}{rm}0.2}}}
\put(6356,-2876){\makebox(0,0)[b]{\smash{\SetFigFont{9}{10.8}{rm}0.1}}}
\put(8529,-2876){\makebox(0,0)[b]{\smash{\SetFigFont{9}{10.8}{rm}0.8}}}
\put(6076,539){\makebox(0,0)[b]{\smash{\SetFigFont{9}{10.8}{rm}events/year}}}
\put(1276,539){\makebox(0,0)[b]{\smash{\SetFigFont{9}{10.8}{rm}events/year}}}
\put(6601,-1561){\makebox(0,0)[lb]{\smash{\SetFigFont{11}{13.2}{rm}NLC}}}
\put(2101,-1561){\makebox(0,0)[lb]{\smash{\SetFigFont{11}{13.2}{rm}LEP}}}
\put(8840,-2876){\makebox(0,0)[b]{\smash{\SetFigFont{9}{10.8}{rm}0.9}}}
\put(9150,-2876){\makebox(0,0)[b]{\smash{\SetFigFont{9}{10.8}{rm}1}}}
\put(2851,-3061){\makebox(0,0)[b]{\smash{\SetFigFont{9}{10.8}{rm}$y_1$}}}
\put(7651,-3061){\makebox(0,0)[b]{\smash{\SetFigFont{9}{10.8}{rm}$y_1$}}}
\put(6046,-2876){\makebox(0,0)[b]{\smash{\SetFigFont{9}{10.8}{rm}0}}}
\put(5972,-2188){\makebox(0,0)[rb]{\smash{\SetFigFont{9}{10.8}{rm}4}}}
\put(5972,-2470){\makebox(0,0)[rb]{\smash{\SetFigFont{9}{10.8}{rm}2}}}
\put(5972,-2752){\makebox(0,0)[rb]{\smash{\SetFigFont{9}{10.8}{rm}0}}}
\put(4350,-2876){\makebox(0,0)[b]{\smash{\SetFigFont{9}{10.8}{rm}0.6}}}
\put(3833,-2876){\makebox(0,0)[b]{\smash{\SetFigFont{9}{10.8}{rm}0.5}}}
\put(3315,-2876){\makebox(0,0)[b]{\smash{\SetFigFont{9}{10.8}{rm}0.4}}}
\put(2798,-2876){\makebox(0,0)[b]{\smash{\SetFigFont{9}{10.8}{rm}0.3}}}
\put(5972,-1905){\makebox(0,0)[rb]{\smash{\SetFigFont{9}{10.8}{rm}6}}}
\put(5972,352){\makebox(0,0)[rb]{\smash{\SetFigFont{9}{10.8}{rm}22}}}
\put(5972, 70){\makebox(0,0)[rb]{\smash{\SetFigFont{9}{10.8}{rm}20}}}
\put(5972,-212){\makebox(0,0)[rb]{\smash{\SetFigFont{9}{10.8}{rm}18}}}
\put(5972,-495){\makebox(0,0)[rb]{\smash{\SetFigFont{9}{10.8}{rm}16}}}
\put(5972,-1623){\makebox(0,0)[rb]{\smash{\SetFigFont{9}{10.8}{rm}8}}}
\put(5972,-1341){\makebox(0,0)[rb]{\smash{\SetFigFont{9}{10.8}{rm}10}}}
\put(5972,-1059){\makebox(0,0)[rb]{\smash{\SetFigFont{9}{10.8}{rm}12}}}
\put(5972,-777){\makebox(0,0)[rb]{\smash{\SetFigFont{9}{10.8}{rm}14}}}
\end{picture}